\documentclass{jfm}
\usepackage{graphicx}
\usepackage{epstopdf, epsfig}

\usepackage{color}
\usepackage{amsmath}
\usepackage{hyphenat}
\usepackage{soul}

\usepackage{float} 

\definecolor{mypink1}{rgb}{0.8392, 0.2745, 0.0353}  
\definecolor{mypink2}{rgb}{0.0745, 0.6235, 1.0000}
\usepackage{hyperref}
\hypersetup{
    colorlinks=true,
    linkcolor=blue,
    citecolor=blue
    }
\usepackage[utf8]{inputenc}
\usepackage[english]{babel}
\usepackage{natbib}
\usepackage{xcolor}
\usepackage{epstopdf, epsfig}
\usepackage{mathabx}
\usepackage{tikz}
\newcommand{\tikzcircle}[2][red,fill=red]{\tikz[baseline=-0.5ex]\draw[#1,radius=#2] (0,0) circle ;}%

\shorttitle{Drop-impact sound} 
\shortauthor{Yang, Tian, Li \& Thoroddsen} 
\title{Sound emission from oscillating bubbles trapped by the collapse of drop-impact craters}
\author{Zi Qiang Yang\aff{1}, 
        Yuan Si Tian\aff{1,2},  
        Er Qiang Li\aff{2},         
 \and Sigur$\eth$ur Tryggvi Thoroddsen\aff{1} 
}

\affiliation{
\aff{1}Division of Physical Sciences and Engineering,
King Abdullah University of Science and Technology (KAUST),
Thuwal, 23955-6900, Saudi Arabia
\aff{2}State Key Laboratory of High Temperature Gas Dynamics, School of Engineering Science, University of Science and Technology of China, Hefei, 230027, China
\aff{}\textbf{Corresponding authors:} 
Zi Qiang Yang, ziqiang.yang@kaust.edu.sa;\\
Sigur$\eth$ur T. Thoroddsen, sigurdur.thoroddsen@kaust.edu.sa
}

\begin{document}
\maketitle
\begin{abstract}
When a drop impacts a deep pool, it forms a crater which subsequently rebounds.
Under certain conditions, a dimple forms at the crater bottom, which pinches off to entrap a small bubble. 
The oscillation of this entrapped bubble is the primary source of the underwater sound produced by rain.
We use simultaneous ultra-high-speed video imaging and synchronized acoustic recording, with an immersed hydrophone, to investigate the details of the sound formation, over a range of impact Weber numbers and different dimple shapes.  With frame-rates as high as 5 million fps, we can track the shape evolution of the pinched off dimple-bubble, which experiences large volumetric compression, by as much as 50\%.  The subsequent volume oscillations are consistent with the observed $\simeq 125$ Pa acoustic pressure amplitude, for the strongest compression.  For our configuration the sound amplitude increases for smaller bubbles pinched off from the dimple.  The acoustic forcing mechanism is therefore the inertial focusing of the momentum of the liquid outside the dimple, as it pinches off.  The acoustic frequency agrees well with the Minnaert theory for freely oscillating spherical bubbles, of the same size. 
The finer details of the acoustic signal reveal an interplay between the larger dimple bubble and the tiny bubble entrapped during the initial contact between the drop and pool. 
For the singular dimple where no bubble is pinched off, the sound generation has a broader range of frequencies, with the tiny bubble oscillating at $\sim 100$ kHz, after being deformed by the rapid vertical retraction of the dimple below the singular jet. 
\end{abstract}
\section{\label{sec:level1}Introduction}
The impact of a drop on a pool is often associated with the familiar {\it plopping} sound.  It is known that this sound is produced by a bubble entrapped into the pool, during the rebounding of the crater in the free surface.  However, as reviewed below, the details of this mechanism have been obscured by the very rapid motions of this entrapment and the lack of time-resolved imaging.  Applying the latest ultra-high-speed imaging to this problem is indeed the focus of our present study.

\cite{Minnaert1933} first considered the sounds of running water and babbling brooks, suggesting that they arise from the oscillations of entrapped bubbles.  He derived a formula for the natural oscillation frequency of an entrapped bubble, in an infinite domain, by balancing the compressibility of the gas within it to the inertia of the surrounding liquid.  For  spherically symmetric small-amplitude oscillations, 
the natural frequency of such bubbles is evaluated using the formula which now bears his name:
\begin{equation}   
f =\frac{1}{2\pi a_{o}}\sqrt{\frac{3\gamma P_{o}}{\rho_{o}}},
\label{Eq_1}
\end{equation}
where $a_{o}$ is the stationary bubble radius, $\gamma$ is the ratio of the specific heats of the air ($\gamma = 1.4$), and $P_{o}$ is the static pressure in the water around the bubble, which is here approximately equal to the atmospheric pressure.  This relation ignores surface tension and viscosity, but a more general derivation can be extracted from Rayleigh-Plesset equations as described in popular texts \citep{Leighton1994, Brennen2014}. 

\cite{strasberg1953,strasberg1956} studied the effect of irregular shaped bubbles, which for example occurs from bubble pinch-off from a nozzle, but found that the oscillation frequency is only slightly modified from the spherical-shaped  oscillation, which shows the highest frequency.  Therefore, one can characterize the sound, in a more general way, by using the bubble volume rather than an arbitrary radius.

\cite{Franz1959splashes} performed the first in-depth study of the sound produced by drop impacts, combining synchronized hydrophone records and film movies.  He connected the key stages of the impact to their signature in the sound signal. 
First, a short pressure pulse is produced by the initial contact of the drop with the pool, followed by a delay during the crater formation, before a more significant sound-wave packet appears after the collapse of the crater.  The crater rebound forms a dimple from which a bubble is entrained.  He also presented canopy formation and bursting of a bubble sitting at the free surface, but
due to the low frame rate of the imaging, $\sim 50$ fps, it was difficult to accurately associate the timing of physical events to the generation of the sound field.

\cite{Pumphrey1989underwater} confirmed Franz’s earlier overall findings on the sound-generation mechanism,
as well as noting that for a given drop diameter the underwater bubble only formed within a certain range of impact velocities, naming this the region of ‘regular entrainment’. 
\cite{Pumphrey1990} studied further the small air bubble entrapped at the bottom of the crater, at low impact velocity. 
\cite{Oguz1990} used their data to derived the bounds of the regular bubble entrapment region in $Fr-We$ domain.

One motivation to study the emission from impact-entrapped bubbles is the source of the underwater sound of rain in the ocean, which is important for the detection of submarines.
The sound energy in the ocean measured by hydrophones under rain has a peak around 14 kHz, which by Minnaert's equation should emerge from air bubbles of diameter 230 $\mu$m, which is consistent with observations 
(\cite{Pumphrey1989underwater,Oguz1990, Pumphrey1990, oguz1991numerical, Prosperetti1993}).

\cite{Prokhorov2011,Prokhorov2012} used a microphone in air and hydrophone in water, as well as a high-speed camera, with frame-rates up to 20 kfps.  However, the impact conditions they studied were quite complex, without the initial dimple and the clear sound emission arising from secondary impact of the drop pinched off from the tip of the Worthington jet impacting on the remnants of the crater.  The strongest sound waves are two orders of magnitude stronger inside the pool (10 Pa) compared to those in the air (0.15 Pa).   

\cite{Thoroddsen2018}, while studying singular jetting, measured directly from ultra-high-speed video-camera clips (at up to 5 million fps) the compression of the central bubble following the pinch-off and the subsequent large volume oscillations, which occurs at frequencies of 23 $\pm$ 2 kHz, which is slightly above the audible range, which reaches $\sim 20$ kHz. 
This was in good agreement with the Minnaert prediction of the frequency for the bubble with a diameter measured to be of approximately 170 $\mu$m.  This bubble compressed, in volume, by more than a factor of two.  The promise of this imaging capability motivated the current study.

\cite{Chicharro2020} studied the acoustic signal from the impact of large water-drops, identifying the importance of secondary droplets for the sound formation.  The secondary drop which pinches off from the tip of the Wothington jet can be quite large, but impacts from a much smaller height than the original drop and can entrap air-pockets when it interacts with the rebounding crater.  This phenomenon may also explain some of the observations in \cite{Gillot2022}.   These entrapments occur outside the regular regime from \cite{Pumphrey1989underwater}. 
The oscillating air body shows sound emission, which grows initially in amplitude, but remains symmetric and resembles the ``heart signal''.  This group has also studied sound from bubbles released from nozzles \citep{Nelli2022,Vazquez2024} and isolated bubbles entrapped by breaking waves \citep{Nelli2025}, which is relevant for characterizing gas transfer from atmosphere into the ocean. 


The underwater sound must propagate through the free interface for our ears to hear the plopping sound in the air.
How this transfer occurs is still being studied.  \cite{Leighton2012} suggested that the airborne sound is simply the underwater sound field propagating through the water-air interface, but no experimental study was conducted to confirm this hypothesis.
The fizz from a glass of carbonated drink is a familiar sound, arising from the popping of numerous bubbles at the free surface. 
\cite{poujol2021} studied this experimentally in the context of the effervescence of a glass of champagne.  The bursting bubbles are of size below the capillary length, in the range of diameters between 200-1000 $\mu$m, while the dominant frequency is $\simeq 50$ kHz, much above the $\sim 10$ kHz expected from Minnaert theory, if the bubbles were oscillating.  
The authors explained the sound based on a collection of drifting Helmholtz resonators capturing the main features of the fizzing sound.  Keep in mind that the capillary overpressure, $2\sigma/R$, of a 200 $\mu$m diameter bubble is only 1.5\% of the atmospheric pressure -- and even smaller for larger bubbles.   


\cite{Phillips2018} took the lead on measuring simultaneously the sound inside and above the pool, using both a hydrophone and a microphone, in addition to a high-speed camera at 30 kfps.  The pressure amplitude is typically two orders of magnitude lower in the air than the pool.
This $\sim 44$ dB attenuation is successfully modeled, using the bubble oscillation transmitted through the induced vertical motion of the pool free surface right above it.  However, their simplified model of these oscillations of the pool surface would not apply to our dimple evolution, as it neglects the rapid upward jet arising above the pinch-off, which controls the free-surface geometry.  \cite{Gillot2020} repeated a similar experiment with video frame-rates up to 40 kfps.  They confirmed the sound amplitudes above and below the free surface, but during the impact of a satellite droplet.  They also highlight the importance of reverberations, by repeating the test in a swimming pool.   

The underlying mechanism for the acoustic signature of the bubble pinched off from the dimple, has been contemplated by many researchers.  \cite{longuet1990analyticmodel} evaluated capillary pressure, shape oscillations away from the spherical shape, inwards momentum of liquid and asymmetry in the spherical velocity thereby generating a second order monopole.  
Estimates of the Laplace and hydrostatic pressure effects show that they represent a minor 10\% contribution to the noise \citep{pumphrey1990oceanic}. 
However, applying analytical flow solutions to a conical crater ignores the highly transient nature of the crater shapes \citep{yang2020}.
\cite{longuet1989} showed that shape mode coupling seems to play a significant role in the dampening of the oscillations of highly distorted bubbles  (this occurs for example when bubbles fragment in turbulence or when it detaches from a nozzle), but a secondary role in the bubble acoustic excitation.

The pinch-off of the crater dimple appears to be similar to the pinch-off of a bubble from a vertical nozzle, even though this geometry is inverted in the vertical direction.  On the other hand, the nozzle pinch-off is better controlled, as it occurs within a quiescent liquid.  This was studied in detail by \cite{Deane2008} and \cite{czerski2010}, who used dual-flash imaging of the pinch-off in addition to the hydrophone measurements.  The 14-$\mu$s-duration flashes are triggered by a laser beam shone through the pinch-off location.  The neck is known to pinch off between two air-cones \citep{Burton2005,Thoroddsen2007}, which they model with a sphere-cone combination to calculate the overall volume-change by the neck motions.  \cite{czerski2010} included measurements to subtract the internal jet rising inside the bubble.  The volume of this jet is similar to that of the disappearing air cone, for the first 200 $\mu$s, but then trails off.  The reduction in bubble volume can lead to overpressure, which can drive the natural ``breathing mode'' of oscillation.  More recently, \cite{Nelli2022} have shown this model to be less accurate for larger bubbles.  

\cite{prokhorov2021,prokhorov2023} used underwater high-speed imaging as well as a hydrophone to measure both the bubble shape and sound.  He attributes the bubble volume oscillation to the sudden change in capillary pressure during the change in interface shape and surface curvature.  He found that the loss of the bubble interface curvature during the bubble pinch-off process induces the abrupt fall in external pressure and an expansion of gas inside the bubble. This promotes the volumetric oscillations of the pinched-off bubble and accompanying acoustic radiation with a positive half-wave signal.  
However, the largest frame-rate of 14 kfps, was insufficient to resolve the time-evolution of the bubble pinch-off dynamics, while the subsequent neck retraction could be tracked.  The neck radius fits a $t^{1/3}$ power-law, but not the $t^{2/3}$ expected for capillary-inertial dynamics \citep{zeff2000}.  Indeed it is expected that the surface tension becomes irrelevant for the last stage of the neck collapse and $t^{0.54}$ can be expected \citep{Keim2006,Burton2005,Thoroddsen2007,yang2020,Blanco-Rodriguez2021}.  It is noteworthy that the measured volume oscillation is only about $\pm 1.5$\% (their figure 3a).

While the source of the sound, produced by drop impacts on pools, has been identified as the entrapped bubble, some open questions still remain. 
The shapes of the bubbles at pinch-off deviate strongly from spherical, which will influence the inertia of the surrounding liquid.  How much will this change the sound frequency? 
The bubble and sound frequency increases when the bubble size decreases, according to the Minnaert formula, which applies when the bubble size is large.  
Does this continue for the smallest bubbles?  What is the acoustic signature associated with bubble shrinkage up to the critical pinch-off moment, as studied by \cite{Thoroddsen2018, yang2020, Tian2023}?
What is the intensity of the emitted sound and what controls this?
What role does the tiny bubble entrapped upon the initial contact of the drop with the pool  \citep{PeckSigurdson1994,Thoroddsen2003} play in the subsequent acoustic wave packet?
Furthermore, the fastest singular jets produced by the crater collapse and described by \cite{zeff2000,Thoroddsen2018} occur without a bubble being pinched off from the dimple.  Can sound be generated for this condition?

To address these open questions, herein we study the sound generation by the collapse of hemispheric drop-impact craters formed by a drop impacting a pool of the same liquid.  
To accurately capture the rapid shape evolution of the bottom dimple, we need frame rates as high as 5 million fps, orders of magnitude larger than has been available in previous studies. 
\newpage

\begin{figure}
\centering
\includegraphics[width=0.84\textwidth]{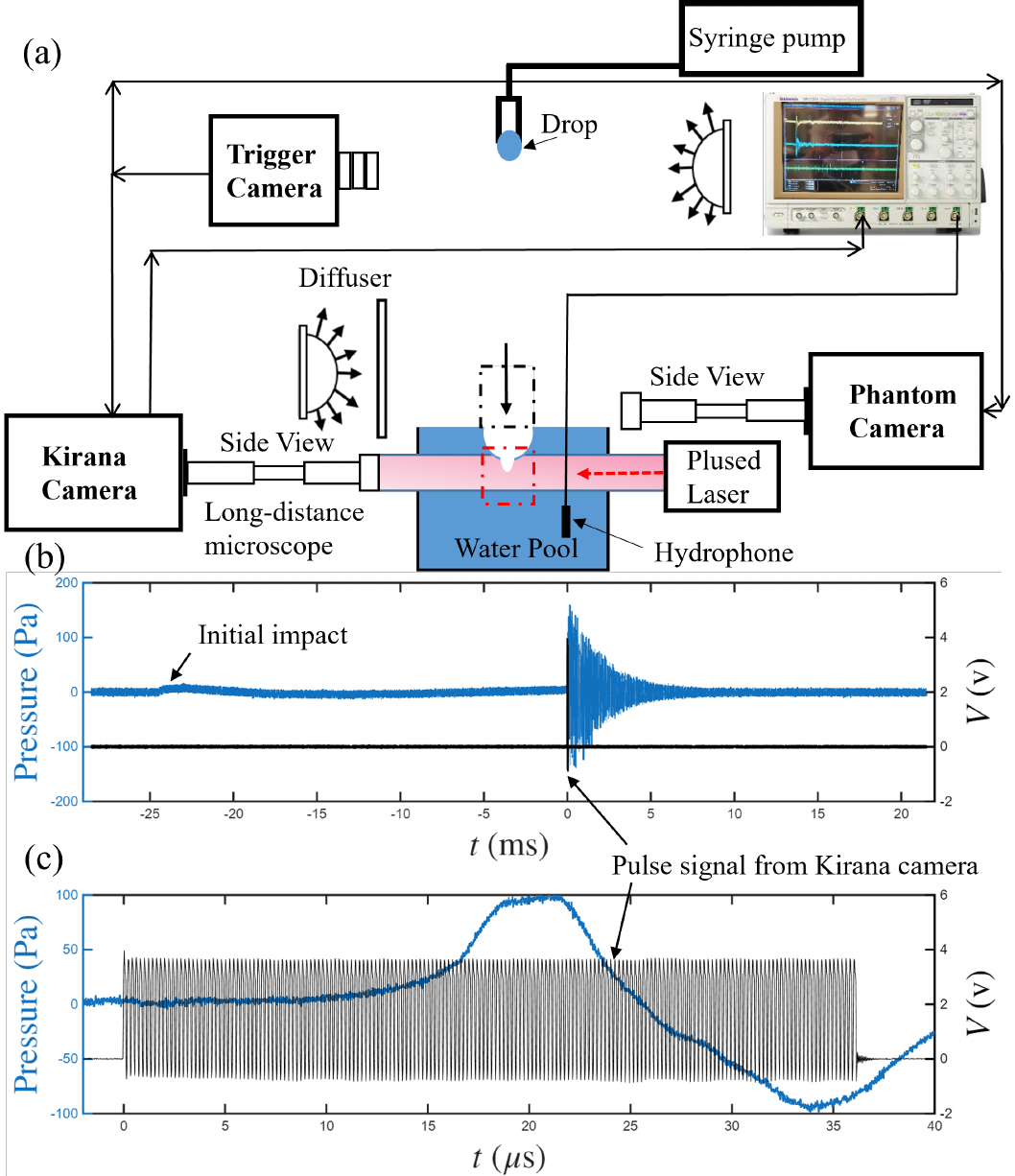}
\caption{
(a) Sketch of the experimental set-up, with two high-speed video cameras viewing the dynamics from perpendicular directions. 
One camera captures the dimple dynamics inside the pool (within red square), while the other views the drop above the pool level to measure the diameter and impact velocity onto its surface.
(b) 
The full long-duration acoustic signature (blue curve) of a drop impact on liquid pool (corresponding to Figure \ref{Fig5_2047_5.0Mfps}), which includes the initial low-amplitude impact pulse, at $t=-25$ ms, caused by the first contact of the drop with the pool surface.  
It also shows the larger-amplitude oscillatory signal from the entrapped bubble. 
The black baseline represents Kirana high-speed camera synchronization signal (Voltage on the right axis).  (c) Zoomed-in details of the 200 ns framing pulses from the Kirana acquisition at 5 Mfps.
}
\label{Fig_01}
\end{figure}

\section{Experimental Setup}
\subsection{Acoustic measurements}
The overall experimental setup is sketched in figure \ref{Fig_01} and is similar to that used in our previous studies, 
which have focused on singular jetting during the crater collapse \citep{yang2020,Tian2023}. 
An inner dimensional 5 cm $\times$ 5 cm $\times$ 5 cm Hellma tank with 2.5 mm thick walls was filled with a water/glycerin mixture.
A small hydrophone (Bruel and Kjaer 8103) was placed at 26 mm depth below the pool surface, oriented approximately $30^o$ off-vertical, \textcolor{black}{angled perpendicular to the direction of the bubbles.}
The hydrophone cable was ensheathed in a water-filled stainless steel tube to decouple it from the acoustic system. 
The drops are slowly pinched off by gravity from a flat stainless steel nozzle with a sharp inner edge of $\simeq$1.2 mm in diameter and fall onto the pool surface.
A syringe pump is used to feed liquid into the drop at a slow flow rate of 10 $\mu$l/min. 
All drop diameters are near 3.6 mm and the liquid tank is sufficiently wide, so capillary waves are not reflected from its walls to influence the impact dynamics.

Two synchronous cameras observe the drop, crater and dimple dynamics. The top camera (Phantom V2511), views from the side above the pool surface, focuses on the drop impact and subsequent jetting, 
while the bottom Kirana camera images the crater and dimple inside the pool, at frame-rates up to 5 Mfps, with $\sim 1\; \mu$m/px resolution when using a long-distance microscope (Leica Z16 APO).  
Back-lighting for the top camera is produced by 350 W Sumita metal-halide lamp shone onto a diffuser, while the total 180 Kirana frames are illuminated by numerous pulsed laser diodes (SI-LUX640, Specialized Imaging) one for each frame.  The pulse-duration of each laser diode is adjustable, but is here set to 100 ns.


\subsection{Triggering mechanism}
The falling drop interrupts an optical line-sensor
(SI-OT3), which is used as a master trigger, sending a signal to the two cameras.
The hydrophone signal is recorded at up to 50 MHz using a Tektronix DPO7254 Digital Oscilloscope.
Simultaneously, the sequence of high-frequency pulses generated with each video frame of the Kirana camera (Specialised Imaging, UK) is sampled on a second oscilloscope channel and is used to freeze the sampling of the acoustic signal from the hydrophone.  
Thereby the video and audio recordings are precisely synchronized, as shown schematically in figure \ref{Fig_01}.

The full acoustic signature of drop impact on liquid pool includes
a clear, but small-amplitude short-duration sound pulse, generated by drop initial impacting onto the liquid surface.  This is followed, $\sim 25$ ms later, by a sudden stronger acoustic pulse followed by a decaying wave-packet from entrapped bubble oscillation, as is shown in figure \ref{Fig_01}(b), by the blue curve.  The conversion of the hydrophone voltage to pressure uses a new factory calibration, repeated after the experiments, as is discussed in Appendix 1.

Figure \ref{Fig_01}(c) shows zoomed-in details of the Kirana high-speed camera synchronization signal with periodic high-amplitude pulses corresponding to the acquisition of each individual video frame.  At the required frame-rate of 5 Mfps drawn in figure \ref{Fig_01}(c), the duration of the 180 frames only spans 36 $\mu$s.  It is therefore clear that even at 1 Mfps, the video imaging only spans a few oscillation cycles of the bubble, while the correspondence of
sound and bubble shape is accurately captured.
\begin{table}
\begin{center}
\def~{\hphantom{0}}
\begin{tabular}{l*{6}{c}r}
Liquid           & Density   & Viscosity & Surface tension & Capillary length \\ 
&    $\rho$ $[kg/m^3]$ & $\mu$ $[mPa.s]$ & $\sigma$ $[mN/m]$ & $L_c$ [mm] \vspace{0.1in}\\ 
W/G mixture & 1140 & 7.3& 68.0& 2.47 \\
\end{tabular}
\caption{Liquid properties of the water/glycerin mixture at 21 $^{\circ}$C.}
\label{tab:1}
\end{center}
\end{table} 
\subsection{Liquids}
The drop and pool are of the same water/glycerin mixture, with viscosity of $\mu=7.3$ cP, which is the same liquid we used to study the crater collapse and singular jets in \cite{Tian2023}.
Distilled water is taken from Milli-Q with internal specific resistance of no less than 18.5 M$\Omega \,$cm$^{-1}$ to confirm its purity. 
At the impact velocities used herein, the air-drag does not deform the drop much, with the variation in the aspect ratio $\beta = D_V / D_H$ being small \textcolor{black}{between 0.94 - 1.10}, deviating only modestly away from the equivalent spherical diameter, $D=(D_H^2 D_V)^{1/3}$, as the drops remain axisymmetric.
Other liquid properties are shown in Table 1, for the room temperature of $T \simeq 21^o$C.  
\subsection{Parameter space}
By varying the drop release-height we produce impact velocities $U$ between 1.27 to 1.50 m/s.
Three dimensionless numbers are used to characterize the drop-impact crater formation: Reynolds, Froude and Weber numbers. 
The values of these parameters, 
fall in the following ranges:
\begin{equation}   
Re = \frac{\rho DU}{\mu} = 693 -815;\hspace{0.15in}
Fr = \frac{U^2}{gD} = 47 - 63;\hspace{0.15in}
We = \frac{\rho DU^2 }{\sigma } = 94 - 130,
\end{equation}
where $g$ is gravity, $\rho$ and $\mu$ are drop density and dynamic viscosity.

\begin{figure}
\centering
\includegraphics[width=0.65\textwidth]{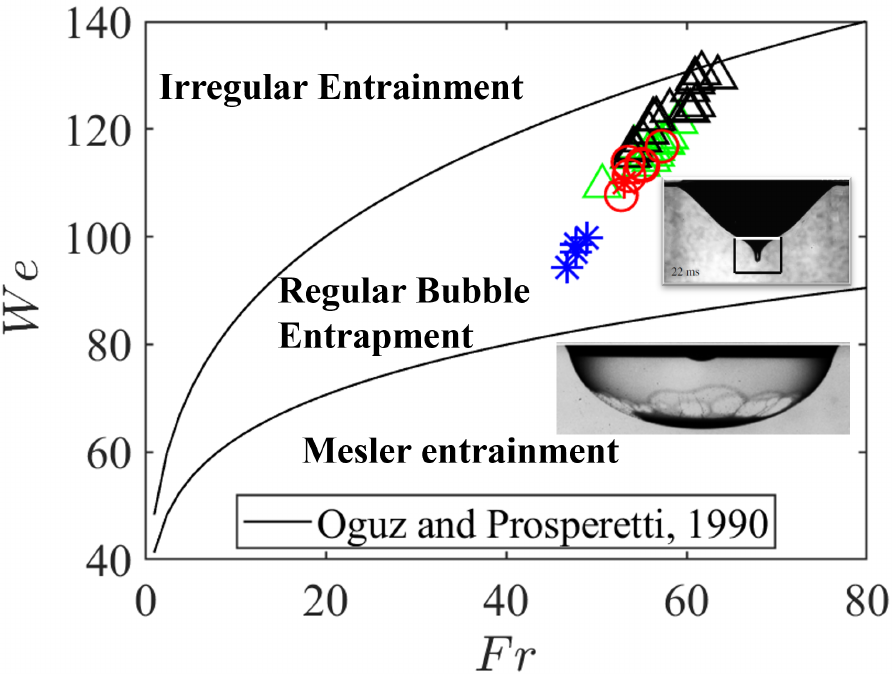}\vspace{-0.01in}
\caption{
Experimental range in the {\it Fr-We} parameter space, for water/glycerin drops impacting a pool of the same liquid.  The two solid curves are the bounds of the regular bubble entrapment regime, where the bottom dimple is pinched off from the rebounding crater, as measured by \cite{Pumphrey1990} and fitted by \cite{Oguz1990}.  
The snapshots of Mesler entrainment and regular bubble entrapment are reproduced from \cite{Thoroddsen2012, Thoroddsen2018}.
The symbols correspond to four typical dimple shapes, with the same names as used in \cite{Thoroddsen2018}.  
Specifically these are: triangular dimple pinch-off (\textcolor{black}{$\triangle$}), teardrop pinch-off (\textcolor{green}{$\triangle$}), small bubble pinch-off (\textcolor{blue}{$*$}) and singular jetting (\textcolor{black}{$\ovoid$}).  The specific shapes for each pinched-off bubble are shown in the next figure.
}
\label{fig2}
\end{figure}
\begin{figure}
\begin{center}
\includegraphics[width=0.85\textwidth]{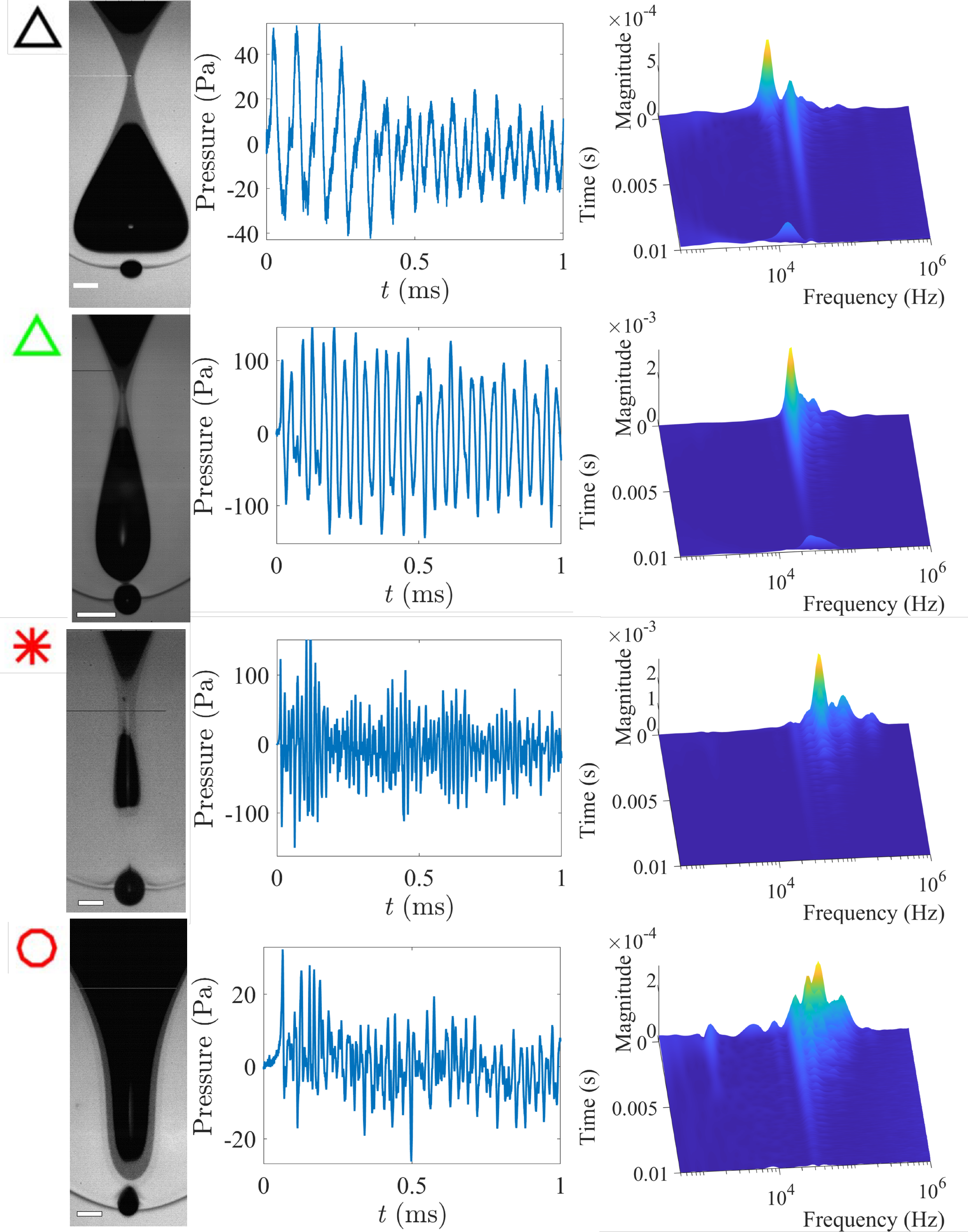}\vspace{0.05in}
\caption{Typical acoustic signals (middle column) and time-frequency wavelet transforms (right side) generated by four different dimple collapse shapes,
resulting from different impact conditions, for the same liquid in the drop and pool. 
These four cases correspond to the main dimple shapes observed in \cite{Thoroddsen2018} over a narrow range of $We$: 
  Triangular dimple pinch-off (\textcolor{black}{$\triangle$}) $D=3.63$ mm, $U=1.41$ m/s, $Fr=56$, $We= 122$; 
  Teardrop pinch-off (\textcolor{green}{$\triangle$}) $D=3.63$ mm, $U=1.34$ m/s, $Fr=51$, $We= 109$;
  Small bubble pinched off (\textcolor{red}{$*$}) $D=3.55$ mm, $U=1.36$ m/s, $Fr=53$, $We= 110$ 
  and singular jet: (\textcolor{red}{$\ovoid$}) $D=3.54$ mm, $U=1.38$ m/s, $Fr=55$, $We= 113$.    
  The scale bars are 100 $\mu$m long. 
  The small bubble visible below the dimples come from the central air-disk entrapped under the drop at its first contact with the pool, see \cite{Thoroddsen2003}.
}
\label{Figure3}
\end{center}
\end{figure}

\section{Results}
\subsection{Acoustic signal for different bubble shapes}
Figure \ref{fig2} shows the parameter regime of the drop impact in $Fr–We$-space, where we have studied the acoustic signals emerging from different dimple morphologies and bubble pinch-off shapes. 
Our impact conditions are compared to the parameter range where the regular bubble entrapment regime occurs, as measured by \cite{Pumphrey1990}.  We use the best-fit curves from \cite{Oguz1990}, which bound this regime and are marked by the solid black lines.
The symbols correspond to four typical dimple shapes  studied by \cite{Thoroddsen2018}, i.e. triangular dimple pinch-off (\textcolor{black}{$\triangle$}), teardrop pinch-off (\textcolor{green}{$\triangle$}), small bubble pinch-off (\textcolor{red}{$*$}) and singular jetting without dimple pinch-off (\textcolor{red}{$\ovoid$}).
Note that because the surface tension and density of the liquid are constant for all experiments herein, the results for different impact velocities, lie along a straight line in the $We-Fr$ diagram.  Keep in mind that the precise value of the singular Weber number is quite sensitive to initial conditions, like the size and shape of the drop.  This sensitivity made \cite{Michon2017}  described the jetting as ``barely reproducible''.

Figure \ref{Figure3} shows an overview of the sound signals emerging from the dynamics of these four different dimple shapes.  The selected video frames show the dimple shapes at pinch-off, or at the start of the vertical retraction for the singular jetting, where the dimple is not pinched off.
The middle column shows the acoustic signals spanning a 1 ms time-duration following the pinch-off, contrasting the amplitudes and frequencies.  The right column shows the amplitude of the wavelet transform of the sound signal.
The tiny bubble visible under all four of the dimples is formed from the thin air-disc entrapped when the drop initially contacts the liquid pool, as was observed by \cite{PeckSigurdson1994} and quantified in \cite{Thoroddsen2003}.
Herein, the liquid has larger viscosity than water and the air-disc contracts into a single bubble and does not form a vertical cylinder which can split into two bubbles \citep{jian2020}.

We first note that the frequency at the peak intensity increases as the pinched-off bubbles become smaller, with the frequency spanning between 10 kHz to 100 kHz.  Furthermore, the signals are not purely of one frequency, showing multiple smaller peaks, while only the teardrop dimple is close to mono-tonal.  From larger time-sequences it is also clear that most of the acoustic energy has dispersed in less than 5 ms, as shown in figure \ref{Fig_01}(b). 

\begin{figure}
\begin{center}
\includegraphics[width=0.80\textwidth]{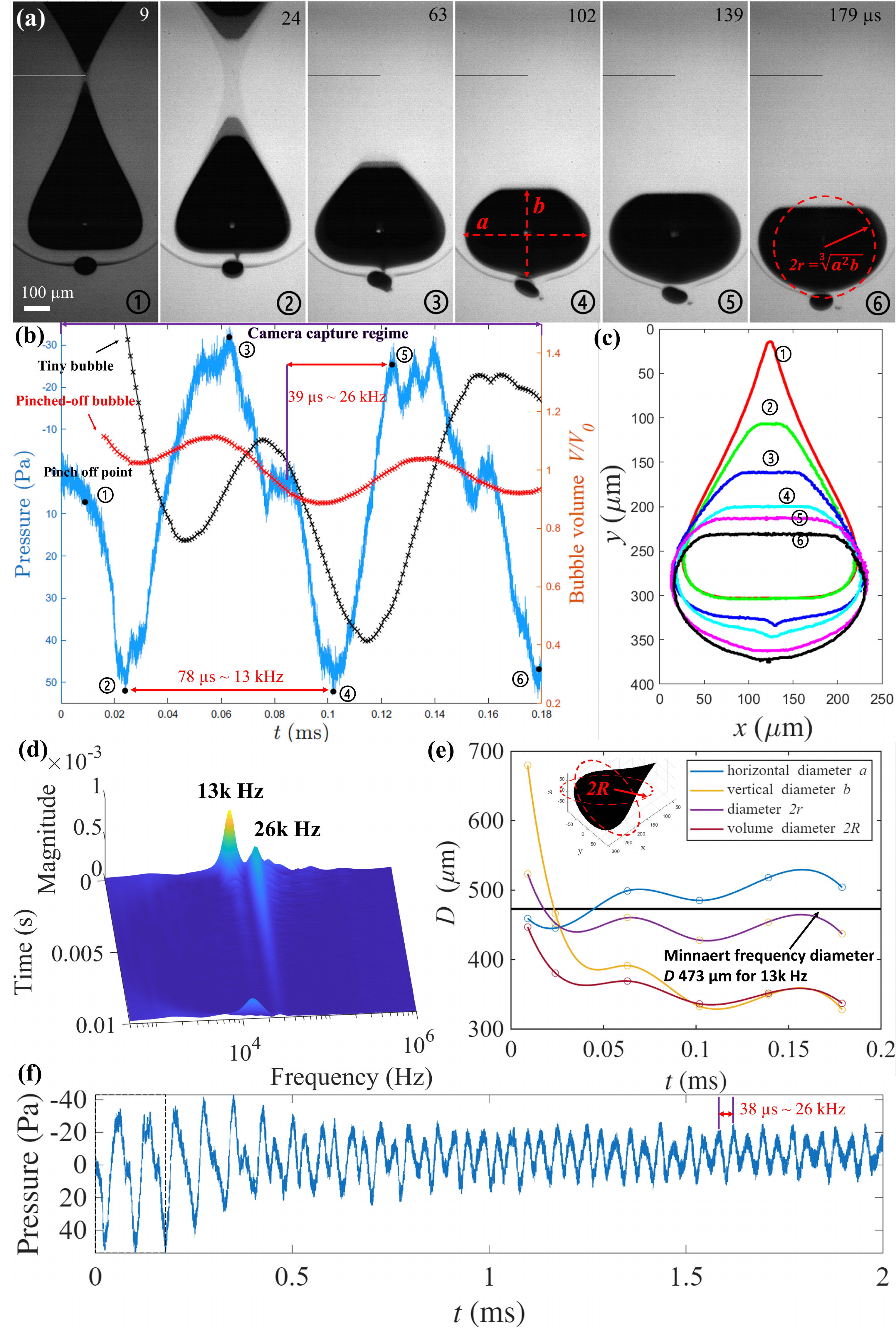}\vspace{-0.1in}\\
\caption{Comparison of high-speed images and acoustical data for the triangular dimple pinch-off (\textcolor{black}{$\triangle$}), for $U$ =  1.41 m/s and $D$ = 3.63 mm, giving $Re$= 802, $Fr$= 56, $We$ = 122. 
(a) Typical frames showing the triangular dimple pinch-off shape and subsequent bubble shape oscillations, taken from a video sequence recorded at 1 Mfps.  The circled numbers marked with \textcolor{black}{$\tikzcircle[fill=black]{1.5pt}$} correspond to the acoustic signal is (b):
\raisebox{.5pt}{\textcircled{\raisebox{-.9pt} {1}}} shows the dimple pinch-off time;
\raisebox{.5pt}{\textcircled{\raisebox{-.9pt} {2}}} The conical neck pulls back and the bubble compresses and reach the minimum volume;
\raisebox{.5pt}{\textcircled{\raisebox{-.9pt} {3}}} bubble expanded to reach maximum volume;
The bubble finishes another compression-expansion cycle, then repeats it in (\raisebox{.5pt}{\textcircled{\raisebox{-.9pt} {4}}}\raisebox{.5pt}{\textcircled{\raisebox{-.9pt}{5}}}\raisebox{.5pt}{\textcircled{\raisebox{-.9pt} {6}}}).
(b) The acoustical signal (left axis) following the dimple pinch-off, including the measured bubble volumes (right axis) from the high-speed video with 1 $\mu$s interframe time.
The normalized volume of the bubble pinched off from the dimple marked with a red line and the tiny bubble from the initial drop impact marked with the dark line. 
(c) The evolution of the bubble shape edge used for entrapped bubble volume calculation with revolution. 
(d) Time-frequency wavelet transform of the sound from dimple pinch-off (\textcolor{black}{$\triangle$}).
(e) The diameter comparison between horizontal and vertical radius, effective radius $r$, $2r= \sqrt[3]{a^{2}b}$, volume radius $R$, and Minnaert prediction.
(f) The damped acoustic signal evolution for extended 2 ms duration.}\vspace{-0.1in}
\label{Figure4_2018_1Mfps}
\end{center}
\end{figure}

\begin{figure}
\begin{center}
\includegraphics[width=0.7\textwidth]{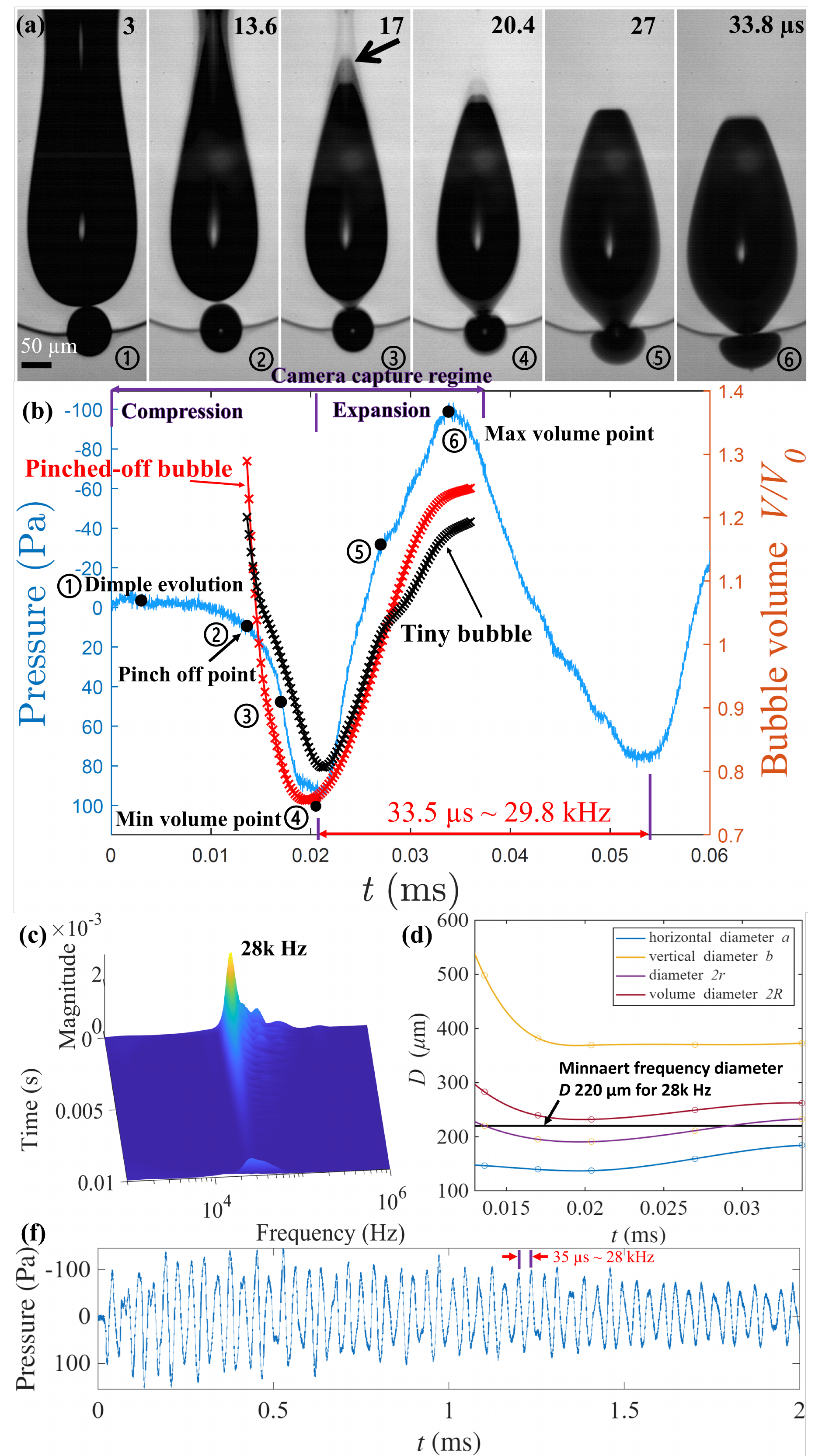} 
\caption{Combined high-speed photography and acoustical data showing the process of teardrop dimple pinch off, for $U$ =  1.34 m/s and $D$ = 3.63 mm, giving $Re$= 760, $Fr$= 51, $We$ = 109.
(a) Typical frames of dimple pinch off and bubble oscillation taken from a video sequence recorded at 5M fps,
corresponding to the circled numbers marked with \textcolor{black}{$\tikzcircle[fill=black]{1.5pt}$} in the acoustic signal in (b);
The scale bar is 50 $\mu$m long.
(b) The acoustical signal (left axis) during the dimple pinch off, with measured bubble volumes (right axis) from high speed camera video.
The normalized volume of bubbles, entrapped bubble pinched off from the dimple marked with red line and tiny bubble from the initial drop impact marked with dark line, are tracked from the dimple pinch off. 
The camera capture regime is also noted with bubble compression and expansion area at the top axis with 200 ns frame time.
(c) Time-frequency wavelet transform of the acoustic signal from teardrop pinch-off (\textcolor{green}{$\triangle$}).
(d) The diameters comparison with Minnaert prediction.
(e) The acoustic signal evolution for 2 ms duration.
}
\label{Fig5_2047_5.0Mfps}
\end{center}
\end{figure}

The following four figures \ref{Figure4_2018_1Mfps} - \ref{Fig7},
show the details connecting the acoustic sound to the dynamics of the above-listed dimple shapes.
This requires very high video frame-rates, for which we use the Kirana camera at up to 5 Mfps.  However, this camera uses in-sensor-image-storage and can only capture a total of 180 frames, which limits the bubble-shape tracking to only the first few oscillations.

\subsection{Triangular dimple}
Figure \ref{Figure4_2018_1Mfps} shows close-up images of a triangular dimple pinch-off and the corresponding acoustic signal, for $We=122$.
The six frames are taken from a Kirana video clip acquired at 1 million fps with spatial resolution of 2.3 $\mu$m/px.  They are selected to show the early bubble-shape evolution.  The times, in $\mu$s, are listed at the top of the frames and marked correspondingly with six black dots, in panel (b), as identified by circled sequential numbers.  The same numbers are used to reveal the evolution of the bubble outline, in figure \ref{Figure4_2018_1Mfps}(c).  In figure \ref{Figure4_2018_1Mfps}(b) the acoustic signal (left axis) marked with the blue line, corresponds to the signal from the top row of figure \ref{Figure3}, but now zoomed in at its start.  Note that the pressure on the $y-$axis has been flipped to ease comparison of the pressure with the bubble volume, i.e. negative pressure is plotted upwards.

The high-speed video allows us to track the changes in the volume of the entrapped bubbles.  The larger bubble entrapped by the dimple pinched-off is marked with a red line in figure \ref{Figure4_2018_1Mfps}(b), while the   tiny bubble from the initial drop contact is marked with the black line.  These volumes are tracked simultaneously, assuming axisymmetry.  The right axis in figure \ref{Figure4_2018_1Mfps}(b) shows the fractional change in the volumes, by dividing by the original reference bubble volumes.

The acoustic pressure increases following the dimple pinch-off at \raisebox{.5pt}{\textcircled{\raisebox{-.9pt} {1}}} and then reaches a maximum at \raisebox{.5pt}{\textcircled{\raisebox{-.9pt} {2}}} coinciding with the compression of entrapped bubble, by the inertial focusing of external liquid.  However, the volume of the triangular bubble only changes by $\sim \pm 10$\%, while the tiny bubble also deforms strongly. 
Then the acoustic signal becomes negative \raisebox{.5pt}{\textcircled{\raisebox{-.9pt} {3}}} coinciding with the expansion of entrapped bubble, considering the spring effect inside the compressible air. 
The larger dimple-bubble finishes one oscillation cycle from compression through expansion into another compression (\raisebox{.5pt}{\textcircled{\raisebox{-.9pt} {1}}}\raisebox{.5pt}{\textcircled{\raisebox{-.9pt}{2}}}\raisebox{.5pt}{\textcircled{\raisebox{-.9pt} {3}}}), then repeats it again in (\raisebox{.5pt}{\textcircled{\raisebox{-.9pt} {4}}}\raisebox{.5pt}{\textcircled{\raisebox{-.9pt}{5}}}\raisebox{.5pt}{\textcircled{\raisebox{-.9pt} {6}}}).

The tiny air-disc bubble undergoes forced oscillation driven by the natural oscillation of larger bubble, but with a 39 $\mu$s time delay, shown in figure \ref{Figure4_2018_1Mfps}(b).  This introduces a phase difference between the two sound waves, which is here about $80^o$. 
Thereby, during each primary acoustic signal oscillation-cycle which takes 78 $\mu$s ($\sim$ 13 kHz), the second acoustic peak is raised at the tiny bubble contraction phase, producing a higher harmonic at twice the primary frequency of $\sim$ 26 kHz, shown in figure \ref{Figure4_2018_1Mfps}(d).
The large bubble oscillation frequency 13 kHz, can also be obtained from its bubble diameter of 473 $\mu$m using Minnaert prediction in Eq. (\ref{Eq_1}). 
The natural Minnaert frequency of tiny bubble (diameter around 70 $\mu$m) would be around 88 kHz, which is not prominent in the wavelet spectrum.  The imaging shows clearly that the volume oscillation of this smaller bubble is purely driven by the pressure field surrounding the larger bubble.  
The two peaks in the acoustic signal, coincide nicely with two peaks in the bubble volumes with the phase offset.
In general the dimple shape at pinch-off is greatly deformed from the spherical, which will affect the validity of the Minnaert prediction of its oscillation frequency.  
For frequency prediction, the effective radius of the entrapped bubble is analyzed using four different approximations of the effective diameter of the bubble, i.e. the horizontal and vertical radius, effective radius $r$,  $2r= \sqrt[3]{a^{2}b}$, and volume radius $R$, measured from the video.  These approximations are shown in figure \ref{Figure4_2018_1Mfps}(e). 
The figure shows that the effective radius $r$ can be used for the best frequency prediction while this may also be the case that the entrapped bubble is not still due to the high frame rate of the video. 
The final stationary bubble radius can be captured for the same impact conditions, but this requires much slower frame rate video, which requires different high-speed camera than the Kirana, which can only acquire 180 frames.  Such slower recordings are used in figure \ref{Fig_09_f_radius}.  Similar comparison is also included for the other dimple shapes in figures 5-7.

\subsection{Teardrop dimple}
Figure \ref{Fig5_2047_5.0Mfps} shows close-up images of the {\it teardrop} pinch-off with the synchronized acoustic signal, for $We=110$, which is slightly below the singular jetting value.
The panel of six frames in figure \ref{Fig5_2047_5.0Mfps}(a) are chosen from a high-speed camera clip taken at 5 Mfps and at optical resolution of $\sim$ 1 $\mu$m/px.  The corresponding times are marked with six dark dots and identified by circled numbers in figure \ref{Fig5_2047_5.0Mfps}(b).
The purple line marked on top of the graph, marks the duration of the Kirana 180 frame video clip which only spans a total of 36 $\mu$s, covering one oscillation cycle of the entrapped bubble. 
This video captures only one cycle of bubble compression and expansion phases.
Comparison with the shapes of the slender dimple, in figure \ref{Fig5_2047_5.0Mfps}(a), show that the pressure is still constant for panel 1, but starts increasing in panel 2, which is still before dimple pinch-off, which occurs between \raisebox{.5pt}{\textcircled{\raisebox{-.9pt} {2}}} and \raisebox{.5pt}{\textcircled{\raisebox{-.9pt} {3}}}, as shown in figure \ref{Fig5_2047_5.0Mfps}(b).
The acoustic signal grows due to compression, reaching maximum at point \raisebox{.5pt}{\textcircled{\raisebox{-.9pt} {4}}}, which coincides with the minima in the bubble volume, as is clearly visible in panel 4 of figure \ref{Fig5_2047_5.0Mfps}(a).
This is followed by the rapid reduction in the pressure signal (keep in mind the sign-change on the y-axis) in \raisebox{.5pt}{\textcircled{\raisebox{-.9pt} {5}}} to reach a maximum volume expansion in \raisebox{.5pt}{\textcircled{\raisebox{-.9pt} {6}}}.
Note the large increase in the bubble width between frames \raisebox{.5pt}{\textcircled{\raisebox{-.9pt} {3}}} and \raisebox{.5pt}{\textcircled{\raisebox{-.9pt} {6}}}.
The duration of one cycle of entrapped bubble is marked by the red bottom line as 33.5 $\mu$s $\sim$ 29 kHz which is close to the peak value of 28 kHz in time-frequency wavelet transform.
The tiny bubble's proximity to the large entrapped bubble results in short time delay, around 3 $\mu$s, for forced oscillation of tiny bubble driven by the large bubble's oscillation shown in figure \ref{Fig5_2047_5.0Mfps}(a \& b).
This phase-locking behavior yields a single distinct frequency peak $\simeq$ 28 kHz in figure \ref{Fig5_2047_5.0Mfps}(c), in clear contrast to the pressure signal for the triangular bubble.

\begin{figure}
\begin{center}
\includegraphics[width=0.76\textwidth]{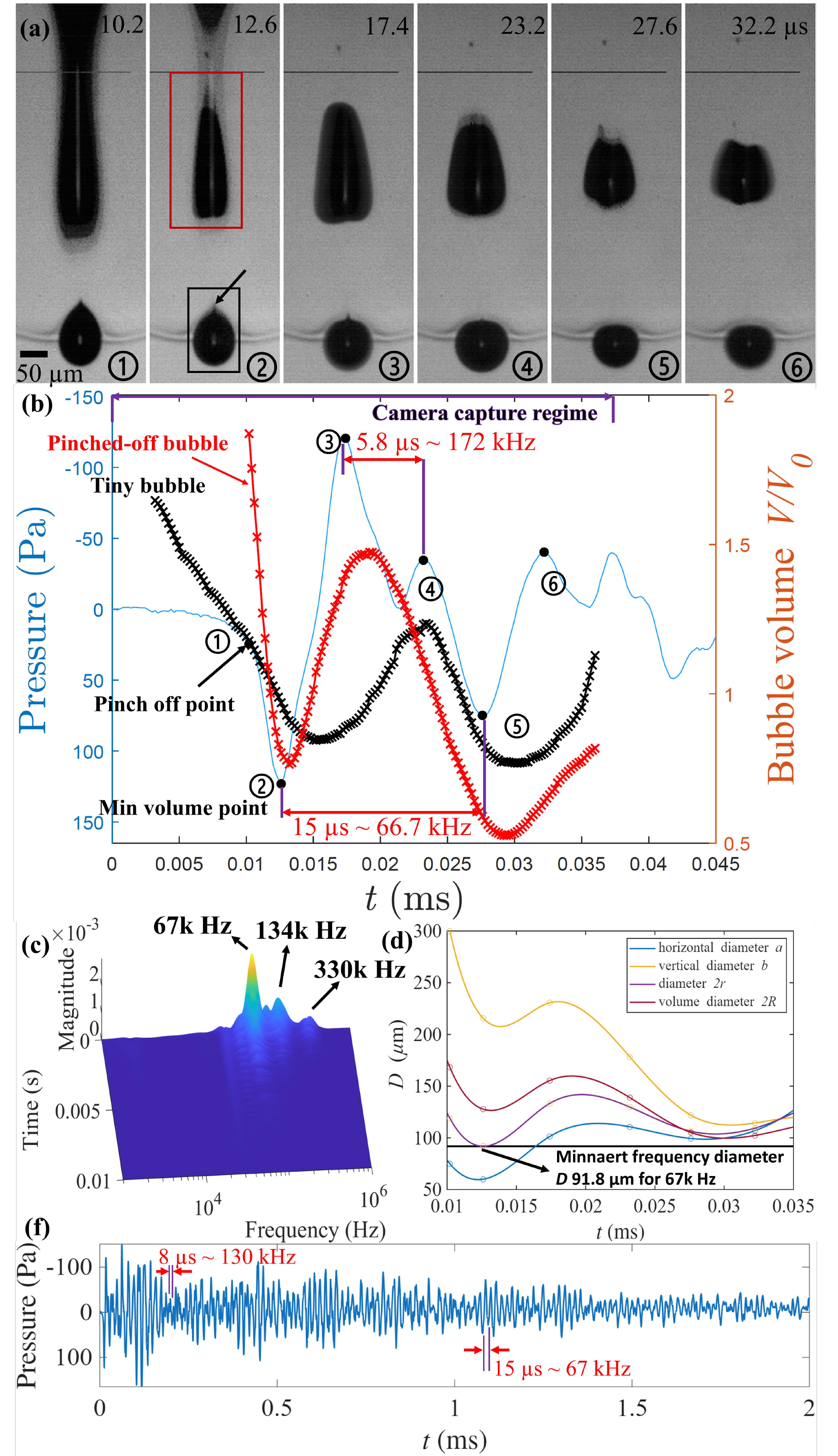}\vspace{-0.1in}\\
\caption{Combined high-speed video and acoustical data showing the evolution of a small-bubble pinch-off, for $U$ =  1.36 m/s and $D$ = 3.55  mm, giving $Re= 754$, $Fr= 53$, $We = 110$.
(a) Typical frames during the dimple pinch-off and bubble oscillation, taken from a video-clip recorded at 5 Mfps. The timing of these frames corresponds to the circled numbers marked with 
black dots on the acoustic signal in (b). Note the large increase in the bubble width between frames \raisebox{.5pt}{\textcircled{\raisebox{-.9pt} {2}}} and \raisebox{.5pt}{\textcircled{\raisebox{-.9pt} {4}}}.
The volume of the pinched-off bubble is marked in red, while the below tiny bubble volume is marked by a black line in (b).
The scale bar is 50 $\mu$m long.
(b) The acoustical signal (left axis) during the dimple pinch-off process with normalized volume of the two bubbles (right axis), measured from the high-speed video.
(c) Time-frequency wavelet transform of the sound from the small-bubble pinch-off (\textcolor{red}{$*$}).
(d) The diameter comparison with the Minnaert prediction.
(e) The acoustic signal evolution for a longer 2 ms duration.
}
\label{Fig6_1041_5M}
\end{center}
\end{figure}

\subsection{Smaller pinched-off bubble}
As the impact Weber number approaches the singular value (at $We \simeq 113$), the size of the bubbles pinched off from the dimple become smaller.  These bubbles experience the largest compression, as they are closer to the {\it ``singularity''} and the converging outer flow has reached smaller radii at time of pinch-off. 
This small-bubble pinch-off process is shown in the zoomed images in figure \ref{Fig6_1041_5M}(a) along with the synchronized acoustics signal in (b).  
The bubble is greatly compressed by $\simeq 50$\% in volume, which is clearly visible in Supplementary Video Clip 3.  This was earlier pointed out for similar impact conditions in \cite{Thoroddsen2018} (their figure 8 and video 7).  Note also that the pressure signal for this case shows the largest observed amplitude of $\pm 125$ Pa.

The bubble pinches off from the dimple between frames 1 and 2.
The video recording is here at 5 Mfps and with spatial resolution of 1.8 $\mu$m/px.
While the strong image ghosting observed again, this can be corrected for by comparison to nearby frames.
It is clearly seen that the bubble continues to be further compressed for a short time, about 2 $\mu$s, after the pinch-off time.  This highlights the role of the momentum of the outer converging liquid and irrelevance of the surface tension.

The tiny bubble is subjected to large compression along with the dimple, at time \raisebox{.5pt}{\textcircled{\raisebox{-.9pt} {2}}}.  
In addition a small cusp emerges from the top of the tiny bubble, marked by the black arrow, induced by the rapid upward motion of the dimple cylinder, driven by the vertical jetting.
The two clear crests in the acoustic signal, noted in figure \ref{Fig6_1041_5M}(b), are synchronized with the maximum volumes of the two entrapped bubbles. 
It is again confirmed that the bubble pinched off from the dimple generates the first and larger acoustic crest, while 
subsequently the forced oscillation of tiny bubble produces the second smaller crest, at a phase-delay of about 80$^o$, both within one cycle of the acoustic signal.
The frequency of the acoustic signal is here estimated from the time of one cycle of acoustic signal, as 15 $\mu$s $\simeq$ 67 kHz and the time-delay of tiny bubble oscillation also induces a higher harmonic with a second time-scale of 
$\simeq 5.8 \;\mu$s, which gives rise to the prominent higher frequency components in the wavelet transform, which is shown in figure \ref{Fig6_1041_5M}(c). This role of the tiny bubble, in shaping the acoustic signal, has not been previously observed.


\begin{figure}
\begin{center}
\includegraphics[width=0.76\textwidth]{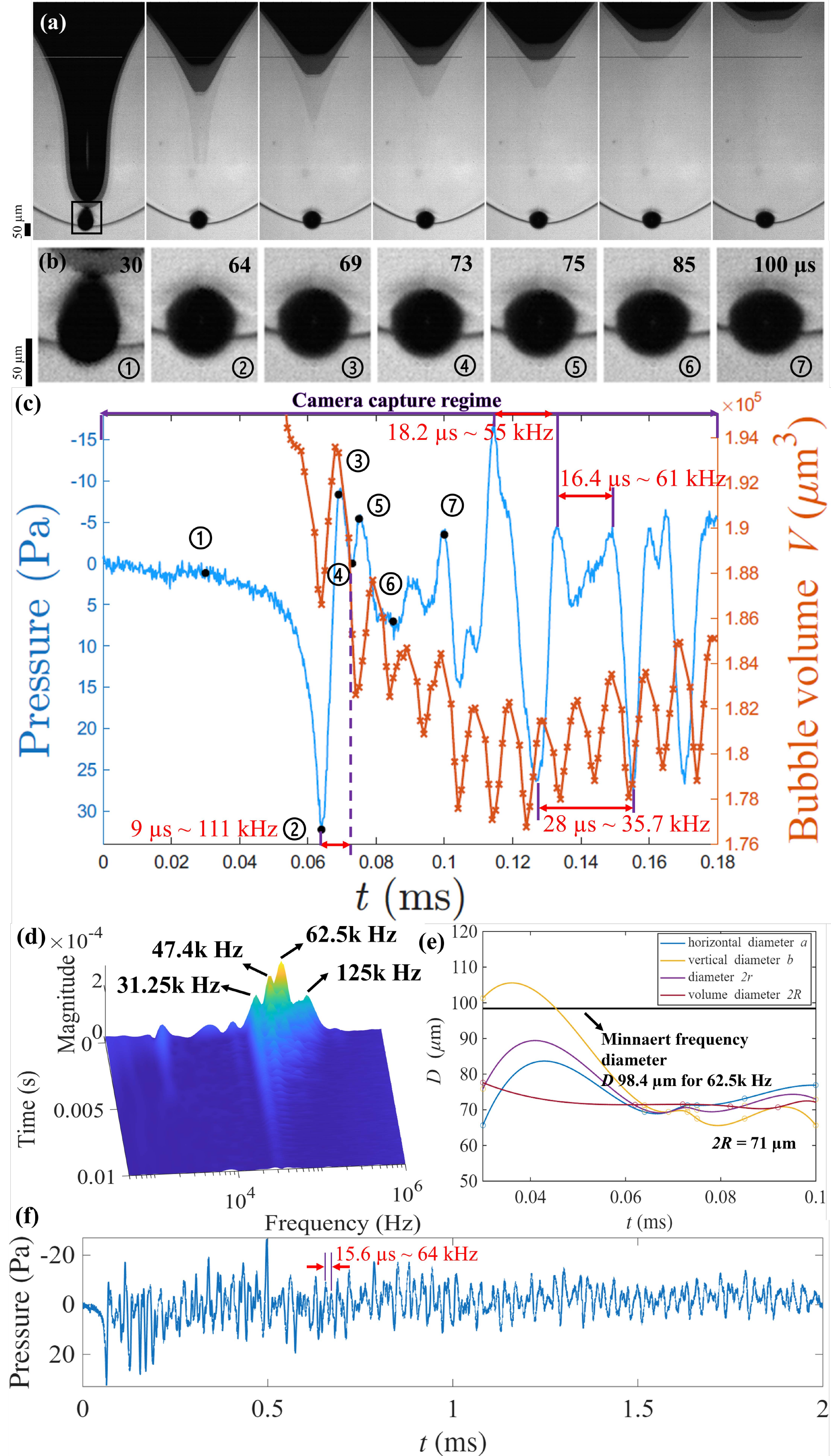}\vspace{-0.1in}\\
\caption{Combined high-speed photography and acoustical data during dimple contraction without bubble pinch-off, for $U$ =  1.38 m/s and $D$ = 3.54 mm, giving $Re= 765$, $Fr= 55$, $We = 113$. 
(a) Typical video frames of dimple retractions without dimple pinch-off and zoomed-in bubble oscillation (b) taken from a video sequence recorded at 1 Mfps.  The frames correspond to the circled numbers marked with \textcolor{black}{$\tikzcircle[fill=black]{1.5pt}$} in acoustic signal of (c);
Both scale bars are 50 $\mu$m long.
(c) The acoustical signal (left axis) for dimple pinch-off, as well as the normalized volume of the tiny bubble (right axis) from the high-speed video.
The camera capture regime is also noted with two purple lines corresponding to two splitted acoustic peaks at the top axis with 200 ns frame time.
(d) Time-frequency wavelet transform of the sound from singular jet (\textcolor{red}{$\ovoid$}).
(e) The diameters comparison with Minnaert prediction.
(f) The acoustic signal evolution for 2 ms duration.
}
\label{Fig7}
\end{center}
\end{figure}

\subsection{Singular dimple}
The {\it singular jetting} during crater collapse is a particularly important case for generation of secondary droplets, as it produces the narrowest and fastest jets, which break up into numerous minute droplets that can easily become airborne.  As such they evaporate and can act as cloud formation nucleii.   This dimple evolution is shown in figure \ref{Fig7}(a).  Here the dimple forms a slender cylindrical air sheet which quickly retracts upwards as a singular liquid jet is shot up along the centreline.  This occurs without a bubble being pinched off from the dimple, as was studied in details in our earlier work \citep{Thoroddsen2018,yang2020,Tian2023}.

Without a primary larger dimple bubble to drive the oscillation pressure field, the tiny bubble must oscillate at its own natural Minnaert frequency.  Zoomed-in images of the tiny bubble in figure \ref{Fig7}(b) show the same pulling of a cusp from its top, as in figure \ref{Fig6_1041_5M}(a).  This is driven by the rapidly retracting bottom of the dimple, in panel \raisebox{.5pt}{\textcircled{\raisebox{-.9pt} {1}}}.
Then the tiny bubble is disturbed by the surrounding transient pressure field from the fast retracting air cylinder.  From the video we determine that it has rather small volume-oscillations, which occur at $\simeq 100$ kHz, which is close to the 87 kHz Minnaert frequency for its size of 75 $\mu$m.  The wavelet transform, of the hydrophone signal, does not capture a peak at this frequency, but rather there are clearly two frequency harmonics at 62.5 and 125 kHz, in figure \ref{Fig7}(d), that likely arise from interaction with the much larger transient pressure variations from the retracting dimple and the singular jetting.   In the following subsection we investigate this discrepancy between the bubble oscillation and the measured acoustics.


\subsection{Shallow vs singular dimple}
\begin{figure}
\begin{center}
\includegraphics[width=0.9\textwidth]{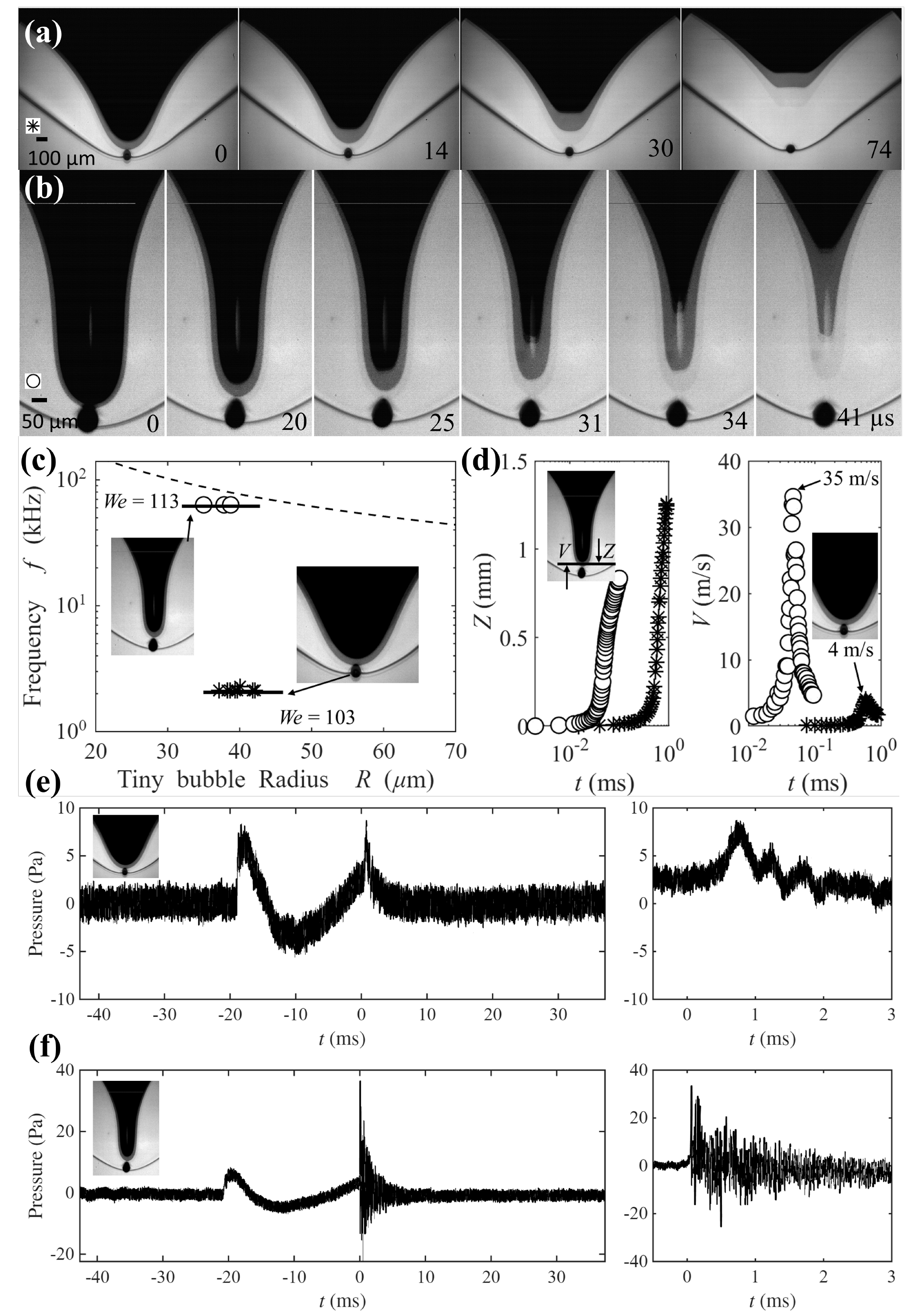}
\caption{Comparison of the acoustics for two differently shaped rebounding dimples, both without bubble pinch-off.
(a) Shallow-angled dimple, for $We= 103$.
(b) Singular dimple rebound, at $We= 113$.
(c) The measured acoustic frequencies for the peak energy, for the two cases.
The dashed line is the prediction of Minnaert’s Equation (\ref{Eq_1}), for a freely oscillation bubble.
(d) Vertical retraction of the bottom tips for the two different dimple shapes, with the open circle for the singular case.
The left panel shows the vertical $Z-$location vs time, while the right panel shows the corresponding retraction velocity of the tips.
(e) \& (f) show the full acoustic pressure signals, corresponding to (a) \& (b) respectively.  The expanded pressure signals, measured near the maximum vertical velocity, are shown on the right-side of the figures.}
\label{Fig_8}
\end{center}
\end{figure}

What causes the discrepancy between the Minnaert frequency and the measured acoustic frequencies for the tiny bubble during the singular jet collapse case?  To investigate this effect, we have compared the singular jetting case to a second rebounding crater, where the bottom angle of the dimple is much shallower, as shown in Fig. \ref{Fig_8}. 
The tiny bubble under the singular dimple oscillates at close to the Minnaert frquency, but this does not show up in the hydrophone pressure signal in Fig. \ref{Fig7}(b).  The acoustic signal for the wider dimple in Fig. \ref{Fig_8}(c) is furthermore, much lower frequency than expected, showing values around 2 kHz, i.e. two orders of magnitude lower than the Minnaert frquencies.  Comparing the two hydrophone signals in Fig. \ref{Fig_8}(e) and (f), it is clear that the amplitude of both signals is very low, at 30 Pa for the singular dimple and only around 5 Pa for the shallow dimple.  The volume change in Fig. \ref{Fig6_1041_5M}(c) is consistent with this, where the volume oscillates by only by $\pm 3$\%.
One must conclude that the dimple rebounding of the shallow dimple ($We \approx 103$) cannot force any significant oscillation of the tiny bubble and is therefore unrelated to the Minnaert mechanism.  The acoustic signal must therefore be promoted by the crater motions.  The lack of narrow energy at the tiny bubble oscillation for the singular case ($We \approx 113$), suggests the rapid singular dimple motion play a dominant role.    
To estimate the role of the dimple retraction we form a second time scale from the retraction velocity, i.e. retraction time-scale, $T_{R} \sim L/V$, 
where the length scale $L$ is the total depth of the retraction air cylinder.  From figures \ref{Fig_8}(a) \& (b) we estimate $L=1.2$ and 0.75 mm, respectively.  The retraction velocity $V$ is 4 m/s for the shallow dimple, while reaching up to 35 m/s for the singular jet, see right panel of figure \ref{Fig_8}(d).  
This gives retraction time-scales of $T_{R} \approx 21\; \mu s$, for the singular case, with the corresponding forcing frequency is $f_{R} = (1/T_R) \approx$ 50 kHz.  On the other hand for the much slower and wider dimple $T_{R} \approx 300\; \mu s$  and $f_{R} = (1/T_R) \approx$ 3.3 kHz. 
These forcing frequencies are in reasonable agreement with the observed values. 

\begin{figure}
\begin{center}
\includegraphics[width=0.75\textwidth]{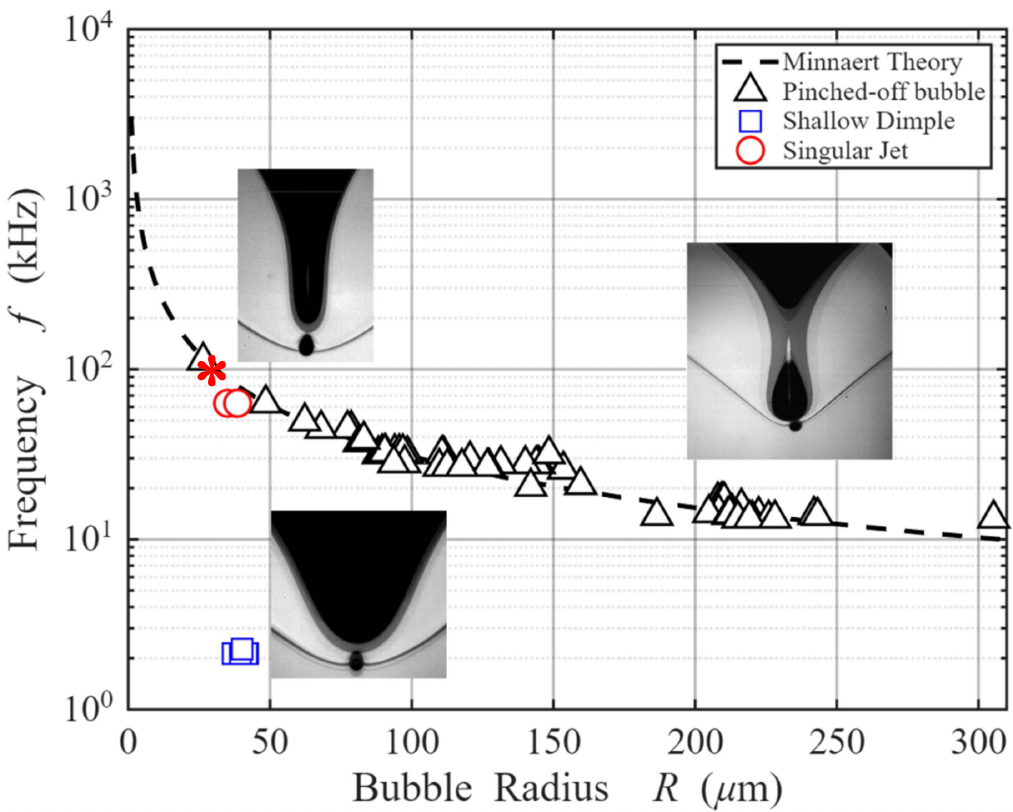}
\caption{The peak acoustics frequency vs size of the bubble entrapped from the dimple pinch-off.
The black dashed line is the locus of Minnaert's equation (\ref{Eq_1}) for a freely oscillating bubble in an infinite domain.
The ambient pressure is taken as atmospheric and the liquid density as that of the water/glycerin mixture.
The ratio of specific heats for the air is $\gamma$ of 1.4 and the bubble is assumed to oscillate adiabatically.
The right inset shows a typical pinched off dimple bubble, while the left insets show two different dimples which do not pinch off a bubble.  The red $*$ shows the frequency of the tiny bubble in figure \ref{Fig7}, determined from the video clip. 
}
\label{Fig_09_f_radius}
\end{center}
\end{figure}

\subsection{Comparison with the Minnaert formulae}
Figure \ref{Fig_09_f_radius} compares the oscillation frequency of the acoustic sound radiated from the various sizes of dimple bubbles to the Minnaert's formulae (\ref{Eq_1}), which is marked by the black dashed line. 
The bubble sizes herein, have diameters from 60 to 600 $\mu$m and are much smaller than in the work by \cite{Pumphrey1990}, who saw diameters between 0.5 - 3 mm.  Our range of diameters is on the other hand slightly broader.  The Minnaert’s Equation (\ref{Eq_1}) still works fine when the diameter of entrapped bubble goes down to 50 $\mu$m.  In this range 
the effect of surface tension and viscosity on Minnaert's equation results in 3\% variation, and can therefore be neglected in the analysis. 

In cases where a dimple does not pinch off, the above subsection shows that the acoustic signature varies significantly with the geometry of the dimple shape. For the shallow broad-angled dimple, its vertical retraction is not strong enough to drive sound from the tiny bubble and the weak amplitude crater-signal produced is predominantly of lower frequency, typically around 2 kHz. In contrast, the formation of a singular jet, a sharper more focused feature, generates a much higher sound frequency, reaching values as high as 63 kHz.  However, high-speed imaging of the volume oscillations of the tiny bubble give a frequency $\simeq 100$ kHz (see figure \ref{Fig7}c), in better agreement with the theory.
\section{Discussion and Conclusions}

\subsection{Acoustic strength}
Knowing the speed and amplitude of the bubble volume oscillations, can one derive the strength of the acoustic emissions? 
Following the book by \cite{Randall2005} (Chapter 3), the basic derivation of sound emission from a three-dimensional axisymmetric sinusoidally oscillating surface, of radius $a$, with a maximum displacement velocity of a fluid element $\xi^{\prime}_m$, gives a solution which has two terms: 
\begin{equation}
\xi^{\prime}_m = B \sqrt{\left( \frac{2\pi}{r\, \lambda}\right)^2 + \left( \frac{1}{r^2}\right)^2 },
\label{Eq_2}
\end{equation}
where $B$ is an integration constant and $\lambda$ is the wavelength of the sound wave. 
The relative strength of these two terms can be directly determined from the video, for example by looking at the images in figure \ref{Fig5_2047_5.0Mfps}.  Here, the bubble oscillation frequency is 28 kHz, so the wavelength is $\lambda = c / f = 1500/28000=54$ mm.  Therefore, at the bubble surface, for the bubble radii of $a=112\; \mu$m, the wavelength of the sound $\lambda >> a$ and the second term above dominates.
Furthermore, between the third and sixth frames in panel (a), we can track the bubble surface moving outwards at $\xi^{\prime}_m \simeq 1.53$ m/s, which we can use, with Eq. (\ref{Eq_2}), to find the constant $B=\xi^{\prime}_m \times a^2= 1.9 $ mm$^3$/s.  The location of the hydrophone is about 20 mm below the dimple, which translates into a displacement velocity of $\xi^{\prime}_m|_{r=0.02m} = 1.9\times 10^{-9}/0.02^2 = 5\; \mu$m/s.  The acoustic pressure can also be calculated, using the liquid density $\rho_o$, as
\[
p_m|_{r=0.02m} = \rho_o \left( \frac{\partial \phi}{\partial t} \right)_m|_{r=0.02m} = \rho_o \left[ \frac{B (2\pi f)}{r} \right]_m|_{r=0.02m} \simeq 61\; Pa,
\]
which is in reasonable agreement with the measured value of 90 Pa.  On the other hand, the estimate for the smaller bubble in figure \ref{Fig6_1041_5M} overestimates the pressure by more than twice.  While these estimates are of the correct order, we must conclude that the shape of the expanding bubble will affect the overall sound field, in more complex ways, than a simple spherically symmetric model.  This could also be better characterized using multiple hydrophones.

\begin{figure}
\centering
\includegraphics[width=0.6
\textwidth]{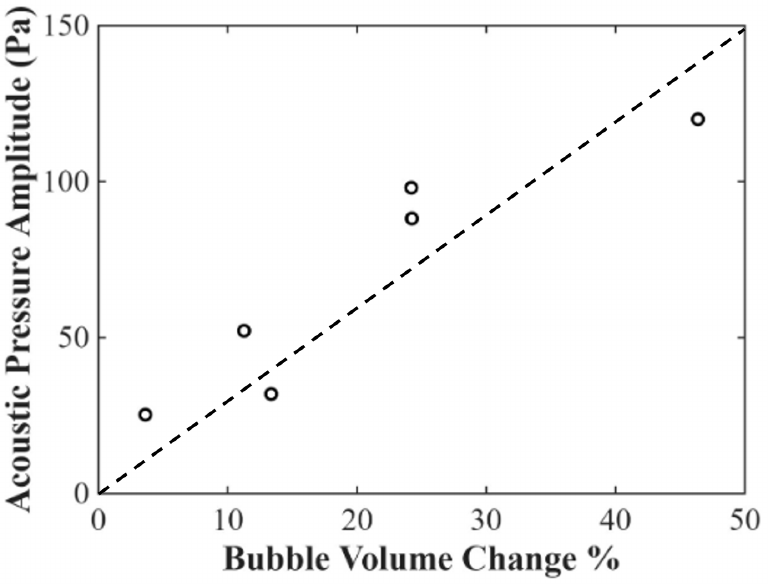}
\caption{\textcolor{black}{The relationship between acoustics pressure amplitude and the maximum compressive change in bubble volume.
The dashed line is added to guide the eye.}
}
\label{Fig_10}
\end{figure}

Figures 4-6 show that compression of the dimple bubbles increases when these bubbles become smaller. 
Investigating figures 3-6 it is clear that the amplitude of the sound pressure signal also increases proportionally to the compression of the pinched-off bubble.  Figure \ref{Fig_10} collects this data revealing this linear relation.  
This can be understood by the convergent flow outside of the dimple, which is driven by a pressure gradient which increases towards the singular point, i.e. smaller dimple stronger compression.   This shows clearly that the acoustic forcing arises from the inertial focusing of the momentum of the outer liquid, compressing the air.

\subsection{Bubbles from nozzles vs crater collapse}
Numerous studies have looked at the sound generated by a bubble pinching off from an underwater vertical nozzle \citep{Deane2008,czerski2010,Nelli2022,Vazquez2024} as a model of the crater sound dynamics.  While this iconic configuration is interesting on its own, we believe it is a poor model of the dimple pinch-off during drop-impact crater collapse.  First, the nozzle pinch-off starts from rest and not subject to strong air-flow through the neck \citep{Gordillo2005,Gekle2010}.  In contrast the dimple pinch-off arises from convergent waves, which travel down the crater wall as is tracked in \cite{yang2020}.  The motions are principally radial for the nozzle, compressing the gas only near the pinch-off location, for the dimple the flow converges also from the bottom and compresses the entire dimple.  This is seen by the volume compression of the dimple, which can reach up to 50\%, while for the nozzle it is more than an order of magnitude smaller.   On the other hand our configuration, with higher liquid viscosity, is more regular than for rain on water and is therefore more singular.  In Appendix B we prove that the pinch-off event itself does not produce strong sound. 
Keep in mind that the final stage of bubble pinch-off events are not universal, as for drop pinch-offs \citep{Day1998}.

\subsection{Summary}
Using an ultra-high-speed video camera synchronized with a hydrophone we have captured the details of sound emission from a bubble entrapped during the rebounding of a drop-impact crater formed in a pool surface.  This allows us to perform time-resolved measurements of the bubble volume over a range of different bubble shapes, which are pinched off from the bottom dimple.  We focus on Weber numbers near those for the singular jetting \citep{Thoroddsen2018,Tian2023}.  Here the inertial focusing of the momentum in the outer liquid leads to strong compression of the air in the dimple.  This compression becomes stronger as one approaches the singularity, i.e. for smaller dimples.  We observe volume compression as large as 50\%.  The subsequent oscillations of the dimple-bubble after pinch-off drive the acoustic pressure waves, with the largest amplitude for the smallest bubble.  In this respect, the sound generation differs fundamentally from that generated by a bubble pinching off from a vertical nozzle.  From the high-speed imaging we infer the role of the tiny bubble, entrapped by the initial contact of the drop with the pool, in giving the double crest of the primary acoustic signal.  This bubble is forced to oscillate as the surrounding pressure field, but at a slight phase delay, owing to its distance.

For the singular jetting dimple, without bubble pinch-off, the exact frequency of the acoustic signal cannot be predicted, but the sound is weak and appears to be driven by the rapid vertical motion of the dimple tip.
In conclusion, for the pinched-off dimple bubble, the resonant frequency in the hydrophone signal agrees well with the prediction by the Minnaert formulae.



\backsection[Acknowledgements]{This study was supported by King Abdullah University of Science and Technology (KAUST) under grant BAS/1/1352‐01‐01. 
E. Q. Li and Y. S. Tian gratefully acknowledge the financial support provided by the National Natural Science Foundation of China (12388101, 12302350).
}

\backsection[Declaration of interests]{ The authors report no conflict of interest.}

\backsection[Author ORCID]{\\
Z. Q. Yang, https://orcid.org/0000-0002-7760-2996;\\
Y. S. Tian, https://orcid.org/0000-0002-9705-2995;\\
E. Q. Li, https://orcid.org/0000-0002-5003-0756;\\
S. T. Thoroddsen, https://orcid.org/0000-0001-6997-4311.}

\begin{appendix}
\begin{figure}
\centering
\includegraphics[width=1.0\textwidth]{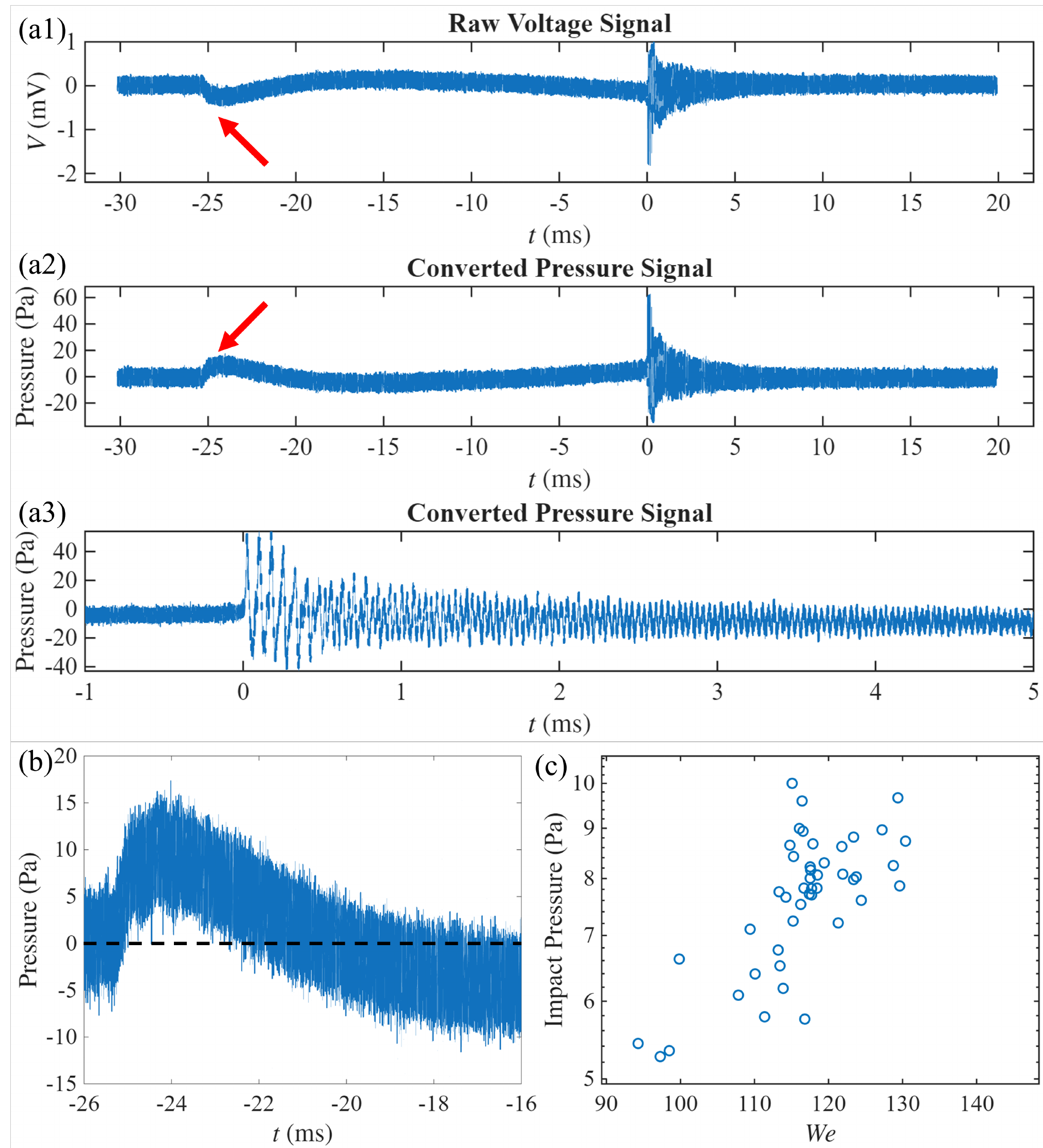}\vspace{0.05in}
\caption{Drop-impact pressure measurement and analysis corresponding to the signal in figure \ref{Figure4_2018_1Mfps}, for $Re= 802$, $Fr= 56$, $We = 122$.
(a1) Raw voltage signal from the hydrophone.
(a2) Pressure signal converted from the raw voltage in (a1) using factory calibration which was repeated after the experimental runs.  Our particular electrical connections gave a negative voltage, which requires changing the sign to obtain the true pressure signal.
(a3) Zoomed-in acoustic signal during the bubble oscillation induced by the pinch-off, converted from the raw voltage in (a1).
(b) Zoomed-in view of the initial drop impact pressure pulse, marked with the red arrow in (a2).  The signal is without noise-level filtering.
(c) Collection of the maximum initial impact pressure, for numerous drop impacts, as a function of the Weber number.
}
\label{Fig_18}
\end{figure}
\section{Pressure from initial drop impact}\label{appB}
The procedure for drop-impact pressure measurement and analysis, corresponding to the signal in figure \ref{Figure4_2018_1Mfps} ($Re= 802$, $Fr= 56$, $We = 122$), is detailed here. The process began with the raw voltage signal from the hydrophone, as shown in figure \ref{Fig_18}(a1). This voltage was then converted into a pressure signal in figure \ref{Fig_18}(a2) by applying the factory calibration, repeated after the experiments. It should be noted that our electrical connections produced a negative voltage, necessitating a sign change to recover the true pressure. The converted signal was then zoomed in to examine the pressure variations during the pinch-off induced bubble oscillation in figure \ref{Fig_18}(a3). Furthermore, the initial drop impact pressure pulse, marked by a red arrow in figure \ref{Fig_18}(a2), is displayed at a higher temporal resolution in figure \ref{Fig_18}(b).  To conclude the analysis, figure \ref{Fig_18}(c) compiles the maximum initial impact pressures from multiple drop impact events, illustrating their dependence on the Weber number, which has the same dependence on impact velocity as the dynamic pressure at the bottom of the drop.

\begin{figure}
\centering
\includegraphics[width=1.0\textwidth]{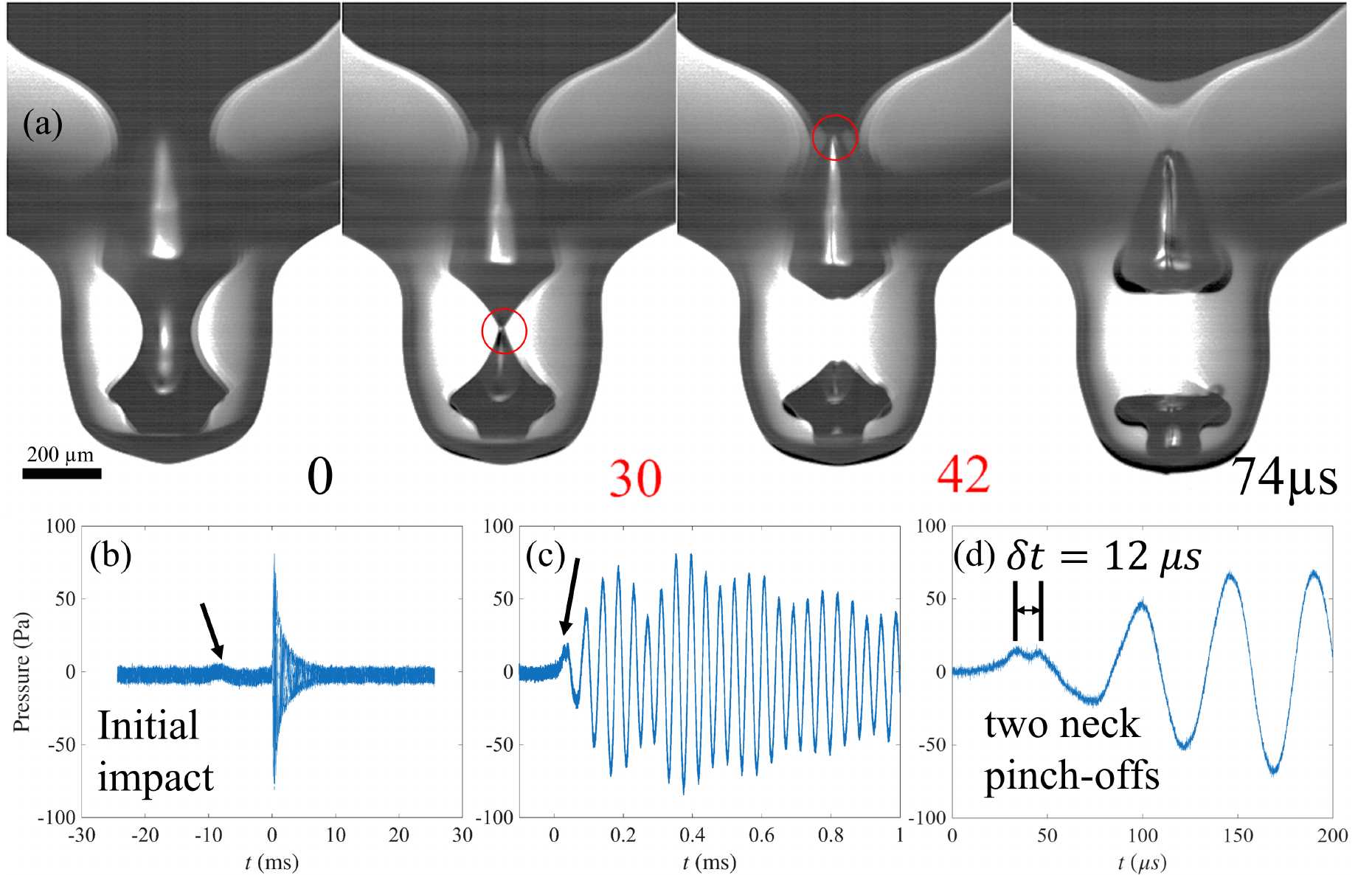}\vspace{0.05in}
\caption{
Double dimple pinch-off from the retracting air dimple following the impact of a PP1 drop onto an immiscible pool of distilled water.  The impact conditions are $U$ = 2.36 m/s, $D$ = 0.73 mm and based on the properties of the PP1 drop, $\rho_{PP1}=1.71$ kg/m$^{3}$, $\mu_{PP1} = 0.81$ cP we get $Re$= 3619, $Fr$= 785 and $We$ = 582. 
(a) Image sequence from high-speed video clip taken at \textcolor{black}{500k fps}.
Red circles indicate the two bubble pinch-off points, which occur 12 $\mu$s apart. The scale bar is 200 $\mu$m long. 
(b) The full acoustic signal recorded by the hydrophone including the initial small positive pressure pulse corresponding to the initial drop contact with the pool surface, followed $\sim 10$ ms later by the bubble oscillation.
(c) Zoomed-in view of the damped wave packet resulting from the oscillations of the bubbles.
(d) Further zoomed-in view revealing the two distinct acoustic pulses generated by the two bubble pinch-off events.
Taken from \cite{yang2022Dissertation}.
See also supplementary movies and a poster submitted to {\it Gallery of Fluid Motion} \citep{Yang2025}.
}
\label{Fig_S1}
\end{figure}

\section{Pinch-off time in acoustic signal}\label{appA}
Considering the pinch-off of the dimple, one might intuitively think that the final collapse of the small air-cylinder would generate strong pressure signal, when the opposite walls collide.  It is therefore of interest to find the precise manifestation of this pinch-off collision in the acoustic signal.  
To accomplish this we build on previous our work on singular jetting for drop-impact craters for immiscible liquids \citep{yang2020}, where a {\it bamboo-shaped} dimple some times appears and presents a  dual pinch-off event.   We again employ high-speed imaging synchronized with hydrophone recordings, as shown in figure \ref{Fig_S1}.
The frames in panel (a) show the two pinch-offs on the retracting air cylinder, first at the bottom neck and almost immediately at the one above it.  
The pinch-off points are marked by red circles and occurring only 12 $\mu$s apart. 
The synchronized hydrophone recording, in figure \ref{Fig_S1}(b), shows the full acoustic signal, where an initial small pulse marked by an arrow corresponds to the drop impact.  This is followed by a slowly dampening modulated wave packet, as shown in figure \ref{Fig_S1}(c). 
Crucially, a further zoomed-in view in figure \ref{Fig_S1}(d) resolves two discrete acoustic pulses that correlate precisely with the two visual pinch-off events. 
This clarifies two aspects of the pinch-off.  First, the pressure signal is positive and of small amplitude, rapidly overtaken by the acoustic waves emerging from the oscillating bubbles.  The modulation is probably generated by the presents of the two bubbles, which are here of complicated shape and contain thin jets visible along the centerline.  The immiscible interface between the two liquids will affect the wave-shape, but should not affect the timing between the two pulses.
This methodology resolves previous ambiguities regarding the acoustic signature generated at the moment of bubble pinch-off in the literature \citep{Deane2008, czerski2010, prokhorov2021, prokhorov2023} and provides insight into the bubble dynamics responsible for underwater noise from rain.


\end{appendix}


%

\bibliography{jfm-references}
\bibliographystyle{jfm}








\end{document}